# Numerical approximation of a phase-field surfactant model with fluid flow

Guangpu Zhu[a], Jisheng Kou[b], Shuyu Sun[c, *], Jun Yao[a], and Aifen Li[a,*]

[a]*Research Center of Multiphase Flow in Porous Media, School of Petroleum Engineering, China University of Petroleum (East China), Qingdao 266580, China*
[b] *School of Mathematics and Statistics, Hubei Engineering University, Xiaogan 432000, Hubei, China.*
[c]*Computational Transport Phenomena Laboratory, Division of Physical Science and Engineering, King Abdullah University of Science and Technology, Thuwal 23955-6900, Kingdom of Saudi Arabia*
\* corresponding authors
Contact information: b16020069@s.upc.edu.cn (Guangpu Zhu), shuyu.sun@kaust.edu.sa (Shuyu Sun)

**Abstract**

Modelling interfacial dynamics with soluble surfactants in a multiphase system is a challenging task. Here, we consider the numerical approximation of a phase-field surfactant model with fluid flow. The nonlinearly coupled model consists of two Cahn-Hilliard-type equations and incompressible Navier-Stokes equation. With the introduction of two auxiliary variables, the governing system is transformed into an equivalent form, which allows the nonlinear potentials to be treated efficiently and semi-explicitly. By certain subtle explicit-implicit treatments to stress and convective terms, we construct first and second-order time marching schemes, which are extremely efficient and easy-to-implement, for the transformed governing system. At each time step, the schemes involve solving only a sequence of linear elliptic equations, and computations of phase-field variables, velocity and pressure are fully decoupled. We further establish a rigorous proof of unconditional energy stability for the first-order scheme. Numerical results in both two and three dimensions are obtained, which demonstrate that the proposed schemes are accurate, efficient and unconditionally energy stable. Using our schemes, we investigate the effect of surfactants on droplet deformation and collision under a shear flow, where the increase of surfactant concentration can enhance droplet deformation and inhibit droplet coalescence.

## 1. Introduction

Surfactants are usually amphiphilic compounds that can absorb as a monomolecular layer to the interface between fluids and reduce the interfacial tension significantly [1-4]. There are two main types of surfactants, insoluble surfactants and soluble surfactants [1]. Insoluble surfactants only absorb onto the interface, while soluble surfactants exist both in the bulk phases and on the interface. The ability of surfactants to control interfacial tension has made them been widely used in may industrial processes, including the manufacture of cosmetics, food processing and oil recovery [5, 6]. Surfactants also play an important role in microsystem with the presence of interface [7, 8], where the capillary effect dominates the inertia of fluids. The presence of surfactants at the interface will significantly alter the dynamical behavior in the microfluidic devices. Thus, it is necessary to understand the behavior of surfactants in a multiphase system. Numerical simulation is taking an increasingly significant position in investigating the interfacial phenomena. However, the computational modelling of interfacial dynamics with soluble surfactant remains a challenging task.

The phase-field model has been used extensively with great successes and has become one of the major tools to resolve the motion of free interfaces between multiple material components [9-12]. Unlike sharp interface models [1, 13-20], the phase-field method utilizes an appropriate free energy functional [21-28], which determines the thermodynamics of a system, to resolve the interfacial dynamics, thus it has a firm physical basis for multiphase flows. In the pioneering work of Laradji et al., the phase-field model was first used to study the dynamics of phase separation of in a binary systems containing surfactants [29]. Since then, a variety of phase-field surfactant models (free energy functional) have been proposed and well investigated. Komura et al. observed some drawbacks of Laradji et al. model and proposed a different two-order-parameter time dependent Ginzburg-Landau model [30]. Theissen and Gompper slightly modified the local coupling term of Laradji et al. model to deal with the same solubility of surfactants in the bulk phases [31]. Samn and Graaf introduced the logarithmic Floy-Huggins potential to restrict the range of local concentration of surfactants [32, 33]. In [34], the authors analyzed the well-posedness of the model proposed by Samn and Graaf, and provided strong evidence that it was mathematically ill-posed for a large set of physically relevant parameters. They made critical changes to the model and substantially increased the domain of validity. This modified model will be used in this study. In [6], the authors introduced a new model that can recover Langmuir and Frumkin adsorption isotherms in equilibrium.

The free energy functional only determines the thermodynamics of a system, and hydrodynamics should also be incorporated into a dynamic multiphase system [32, 35-37]. Numerically, it is extremely difficult to develop unconditionally energy stable schemes [2, 3, 33, 38] for phase-field surfactant models of two-phase incompressible flow. The main difficulties [3, 12] include: (1) the strong nonlinear couplings between multiple phase field variables; (2) the coupling between velocity and multiple phase-field variables through convection terms and nonlinear stresses; (3) the stiffness issue form the interfacial thickness; (4) the coupling of the velocity and pressure through incompressibility constraint. Several attempts have been made to solve the multiphase system with surfactants, including the work of Samn and Graaf [32], Engblom et al. [34], Yun et al. [39], Garcke et al. [35], Teigen et al. [37], Liu and Zhang [6]. Although they have obtained some interesting results using different numerical methods, none of them can provide the energy stability for numerical schemes in theory. It has been observed that numerical schemes with energy instability may introduce an excessive amount of numerical dissipation near singularities, which in turn lead to large numerical errors, particularly for long time integration [12]. Therefore, the main purpose of this study is to construct efficient schemes with unconditional energy stability for the phase-field surfactant model with fluid flow. The schemes should be accurate, totally decoupled and easy-to-implement (only need to solve a sequence of linear, elliptic equations). To develop such schemes, the invariant energy quadratization (IEQ) method is adopted in this study. The IEQ approach [3, 40, 41] introduces a set of new variables to transform the free energy density into an invariant quadratic functional. The new transformed system is exactly equivalent to the original system, and all nonlinear terms can be treated semi-explicitly, which in turn produces a linear system. To the best of the author's knowledge, the proposed schemes here are the first linear, decoupled and energy stable schemes for the hydrodynamics coupled phase-field surfactant model.

The rest of this paper is organized as follows. In Section 2, we describe a nonlinearly coupled phase-field surfactant model with fluid flow. In Section 3, the governing system are transformed into a new equivalent system by using the IEQ approach. Then we construct an efficient, decoupled and energy stable scheme for the transformed system and carry out the energy estimates for the proposed scheme. Several two-dimensional (2D) and three-dimensional (3D) numerical examples are investigated in Section 4 and the paper is finally concluded in Section 5.

## 2. The governing equation

For a multiphase system with soluble surfactants, we consider a typical binary fluid-surfactant phase field model introduced in [34]. The dimensionless free energy of system $E_{ps}$ reads:

$$E_f(\phi,\psi) = E_1(\phi) + E_2(\psi) + \int_\Omega \left( \frac{\psi\phi^2}{4\text{Ex}} - \frac{\psi(\phi^2-1)^2}{4} \right) d\Omega. \quad (1)$$

where $E_1(\phi)$ is the Ginzburg-Landau type of Helmholtz free energy functional [42-44]. The phase-field variable $\phi$ is used to label the volume fraction of two phases ($\phi$ equals to $\pm 1$ in the two homogeneous equilibrium phases, and it varies continuously across the interface between fluids). Cn is the Cahn number denoting the interfacial thickness.

$$E_1(\phi) = \int_\Omega \left( \frac{(\phi^2-1)^2}{4} + \frac{\text{Cn}^2}{4}|\nabla\phi|^2 \right) d\Omega,$$

The first term in $E_1(\phi)$, the double well potential, promotes the two-phase separation. The square-gradient term contributes to the mixing of two phases. The competition between the two terms creates a thin smooth transition layer between two homogeneous equilibrium phases [12].

To represent the surfactant concentration in a multiphase system, another phase-field variable $\psi$ is introduced and the relative free energy reads:

$$\begin{cases} E_2(\psi) = \int_\Omega \text{Pi}\, G(\psi) d\Omega, \\ G(\psi) = \psi\log\psi + (1-\psi)\log(1-\psi), \end{cases}$$

where Pi is the temperature-dependent constant taking the role of a diffusion coefficient for $\psi$. $G(\psi)$ is the logarithmic Flory-Huggins type energy potential restricting the value of $\psi$ to be inside the domain of (0, 1).

The local coupling term $\psi\phi^2/(4\text{Ex})$ in the equation (1) penalizes free surfactants in the bulk phases and stabilizes the phase-field model [34]. The positive parameter Ex controls the bulk solubility. The special molecular composition of surfactants enables them to selectively absorb on the interface, and form a buffer zone to reduce the system energy [6]. The surface energy $-\psi(\phi^2-1)^2/4$ accounts for the high concentration of surfactants near the interface [34].

Chemical potentials $w_\psi$ and $w_\phi$ can be obtained from the variation of free energy functional $E_f$ with respect to two phase-field variables $\psi$ and $\phi$, respectively, that is

$$\delta E_f(\phi,\psi) = \int (w_\psi\, \delta\psi) d\Omega + \int (w_\phi\, \delta\phi) d\Omega.$$

where

$$w_\psi = \text{Pi}\log\left(\frac{\psi}{1-\psi}\right) + \frac{1}{4\text{Ex}}\phi^2 - \frac{(\phi^2-1)^2}{4}, \quad (2)$$

and

$$w_\phi = -\phi + \phi^3 - \frac{\text{Cn}^2}{2}\Delta\phi + \frac{1}{2\text{Ex}}\psi\phi - \psi\phi(\phi^2-1). \quad (3)$$

Minimizing the total free energy $E_f$ with respect to phase-field variables yields the equilibrium condition, where both chemical potentials are constants throughout the system. Deviations from the equilibrium condition, measured by gradient of chemical potentials, lead to diffusive fluxes in the bulk phases, that is

$$\mathbf{J}_\psi = -\frac{M_\psi}{\mathrm{Pe}_\psi}\nabla w_\psi, \quad \mathbf{J}_\phi = -\frac{1}{\mathrm{Pe}_\phi}\nabla w_\phi.$$

The conservation of phase-field variables $\phi$ and $\psi$ means that diffusive fluxes and the material derivatives of phase-field variables satisfy the continuity equation

$$\psi_t + \nabla \cdot (\mathbf{u}\psi) = -\nabla \cdot \mathbf{J}_\psi = \frac{1}{\mathrm{Pe}_\psi}\nabla \cdot M_\psi \nabla w_\psi, \tag{4}$$

$$\phi_t + \nabla \cdot (\mathbf{u}\phi) = -\nabla \cdot \mathbf{J}_\phi = \frac{1}{\mathrm{Pe}_\phi}\Delta w_\phi, \tag{5}$$

where $\mathbf{u}$ is the velocity, $\mathrm{Pe}_\psi$ and $\mathrm{Pe}_\phi$ are Péclet numbers. As in [34], a degenerate mobility $M_\psi = \psi(1-\psi)$, which vanishes at extreme points $\psi=0$ and $\psi=1$, is introduced to combine with the logarithmic chemical potential $w_\psi$. Equations (4) and (5) are typical Cahn-Hilliard-type equations. Equations (2) – (5) are coupled to the hydrodynamic equations in the form [26, 45]

$$\mathbf{u}_t + \mathbf{u}\cdot\nabla\mathbf{u} - \frac{1}{\mathrm{Re}}\Delta\mathbf{u} + \nabla p + \frac{1}{\mathrm{We}}\left(\phi\nabla w_\phi + \psi\nabla w_\psi\right) = 0, \tag{6}$$

$$\nabla \cdot \mathbf{u} = 0, \tag{7}$$

where We = ReCaCn. Re is the Reynolds number, Ca is the Capillary number and $p$ is the pressure. For the domain $\Omega$, periodic boundary conditions or the following boundary conditions can be used to close the system:

$$\partial_\mathbf{n}\phi = \partial_\mathbf{n}\psi = \nabla w_\phi \cdot \mathbf{n} = \nabla w_\psi \cdot \mathbf{n} = \mathbf{u} = \partial_\mathbf{n}p = 0, \text{ on } \Gamma. \tag{8}$$

Equations (2) – (8) form the complete hydrodynamics coupled phase-field surfactant model. The total energy $E_{tot}$ of the hydrodynamics system is a sum of kinetic energy $E_k$ together with the free energy $E_f$:

$$E_{tot} = E_k + E_f = \int_\Omega \left(\frac{\mathrm{We}}{2}|\mathbf{u}|^2 + \frac{(\phi^2-1)^2}{4} + \frac{\mathrm{Cn}^2}{4}|\nabla\phi|^2 + \mathrm{Pi}\,G(\psi) + \frac{\psi\phi^2}{4\mathrm{Ex}} - \frac{\psi(\phi^2-1)^2}{4}\right)d\Omega. \tag{9}$$

We can easily derive the following PDE energy dissipation law for the governing equation (2) – (8):

$$\frac{d}{dt}E_{tot}(\phi,\psi) = -\frac{1}{\mathrm{Pe}_\phi}\int|\nabla w_\phi|^2 d\Omega - \frac{1}{\mathrm{Pe}_\psi}\int\left|\sqrt{M_\psi}\nabla w_\psi\right|^2 d\Omega - \mathrm{CaCn}\int|\nabla\mathbf{u}|^2 d\Omega \leq 0. \tag{10}$$

Now we analyze the equilibrium profile for surfactant concentration $\psi$, which is important for subsequent numerical simulations [34]. Considering the fact that chemical potential $w_\psi$ is constant in the equilibrium, we have $w_{\psi_b} = w_{\psi(x)}$

$$w_{\psi_b} = \mathrm{Pi}\log\left(\frac{\psi_b}{1-\psi_b}\right) + \frac{1}{4\mathrm{Ex}}\phi_b^2 - \frac{(\phi_b^2-1)^2}{4}, \tag{11}$$

$$w_{\psi(x)} = \mathrm{Pi}\log\left(\frac{\psi(x)}{1-\psi(x)}\right) + \frac{1}{4\mathrm{Ex}}\phi^2 - \frac{(\phi^2-1)^2}{4}, \tag{12}$$

where $\psi_b$ is the surfactant concentration in the bulk phases $\psi(\infty)$, and $\phi_b$ is the bulk value $\phi(\infty)$. Subtracting (11) from (12), and introducing the intermediate variable $\psi_c$, we get the relation:

$$\text{Pi}\log\psi_c = -\frac{1}{4\text{Ex}}\left(\phi_b^2 - \phi^2\right) - \frac{1}{4}\left[\left(\phi_b^2 - \phi^2\right)\left(2 - \phi_b^2 - \phi^2\right)\right]. \tag{13}$$

Then the equilibrium profile for $\psi$ can be obtained

$$\psi(x) = \frac{\psi_b}{\psi_b + \psi_c(1-\psi_b)}. \tag{14}$$

Since the surfactant bulk concentration $\psi_b$ is small ($\psi_b \ll 1$), $\phi = 0$ on the interface of fluids and $\phi_b = \pm 1$, equations (14) and (13) can be simplified to [34]

$$\psi_0 = \frac{\psi_b}{\psi_b + \psi_c}, \tag{15}$$

$$\text{Pi}\log\psi_c = -\frac{1}{4}\left(1 + \frac{1}{\text{Ex}}\right), \tag{16}$$

where $\psi_0$ is the equilibrium concentration of surfactants on the interface. Equation (15) is the typical Langmuir isotherm and $\psi_c$ is the Langmuir adsorption constant. Given specific Pi and $\psi_c$, Ex can be obtained from equation (16).

## 3. Numerical schemes

### 3.1. Transformed governing system and its energy law

We now develop an energy-based diffuse interface model and construct some efficient time marching schemes to solve the nonlinearly coupled phase-field surfactant model with fluid flow. The desired schemes should be accurate, totally decoupled, unconditionally energy stable and easy-to-implement. From the numerical point of view, it is quite a challenging topic to construct such schemes. The main difficulties include (1) the strong nonlinear couplings between phase-field variables; (2) the treatment of nonlinear potentials, such as the Ginzburg-Landau double well potential and Flory-Huggins potential; (3) the nonlinear coupling terms between phase-field variables and velocity through stress and convective terms; (4) the coupling of velocity and pressure through the incompressibility constraint.

For the first two difficulties (1) and (2), the commonly used techniques, such as convex splitting and stabilization approaches, may not be optimal choices due to some imperfections [3, 46]. The convex splitting approach [47-49] normally produces nonlinear schemes, thus the implementations are often complicated and the computational costs are high [46]. Schemes constructed by the stabilization approach [12, 25] are usually linear, energy stable and easy-to-implement. However, its stability requires that the classical PDE solution and the numerical solution hold the maximum principle, which is very hard to prove [46]. Therefore, a novel IEQ approach in this study is adopted to deal with nonlinear potentials, and it has been successfully applied to solve a large class of gradient flows [2, 3, 40, 41, 50]. Most recently, the authors have applied the IEQ approach to the case of multiple nonlinearly coupled variables in [2, 3]. For the difficulty (3), the computations of phase-field variables and velocity are decoupled through a subtle implicit-explicit treatment. For the difficulty (4), we use a classical projection method to decouple the computation of velocity and pressure.

Before using the IEQ approach to transform the governing system, we first regularize the logarithmic potential $G(\psi)$ from the domain (0, 1) to (-∞, +∞). For any $\xi > 0$, the regularized logarithmic potential [3, 51] can be written as

$$\hat{G}(\psi) = \begin{cases} \psi \ln \psi + \dfrac{(1-\psi)^2}{2\xi} + (1-\psi)\ln\xi - \dfrac{\xi}{2}, & \text{if } \psi \geq 1-\xi, \\ \psi \ln \psi + (1-\psi)\ln(1-\psi), & \text{if } \xi \leq \psi \leq 1-\xi, \\ (1-\psi)\ln(1-\psi) + \dfrac{\psi^2}{2\xi} + \psi \ln\xi - \dfrac{\xi}{2}, & \text{if } \psi \leq \xi. \end{cases} \quad (17)$$

When $\xi \to 0$, $\hat{G}(\psi) \to G(\psi)$, and we consider the numerical solution to the model formulated with the regularized function $\hat{G}(\psi)$. The ˆ is omitted in the notation for convenience.

The IEQ approach introduces a set of new variables to transform nonlinear potentials into invariant quadratic forms. The new transformed system is exactly equivalent to the original system, and all nonlinear terms can be treated semi-explicitly. More precisely, we define two auxiliary variables

$$U = \phi^2 - 1, \quad V = \sqrt{G(\psi) + B},$$

where $B$ is a positive constant to ensure $G(\psi)+B>0$, and $B = 1$ is adopted in this study [3]. Then the total free energy functional $E_{tot}$ can be rewritten as

$$E_{tot}(\phi, \psi, U, V) = \int \left( \frac{\text{We}}{2}|\mathbf{u}|^2 + \frac{U^2}{4} + \frac{\text{Cn}^2}{4}|\nabla\phi|^2 + \text{Pi}V^2 + \frac{\psi\phi^2}{4\text{Ex}} - \frac{\psi U^2}{4} \right) d\Omega - \text{Pi}B|\Omega|,$$

Note that the total free energy functional $E_{tot}$ is not changed due to the introduction of the zero term P$iB$–P$iB$. With the reformulated free energy functional, we can obtain a new and equivalent governing system [26, 45]

$$\psi_t + \nabla \cdot (\mathbf{u}\psi) = \frac{1}{\text{Pe}_\psi} \nabla \cdot M_\psi \nabla w_\psi, \tag{18a}$$

$$w_\psi = \text{Pi}H(\psi)V + \frac{\phi^2}{4\text{Ex}} - \frac{U^2}{4}, \tag{18b}$$

$$V_t = \frac{H(\psi)}{2}\psi_t, \tag{18c}$$

$$\phi_t + \nabla \cdot (\mathbf{u}\phi) = \frac{1}{\text{Pe}_\phi} \Delta w_\phi, \tag{18d}$$

$$w_\phi = -\frac{\text{Cn}^2}{2}\Delta\phi + \phi U + \frac{\psi\phi}{2\text{Ex}} - \psi U \phi, \tag{18e}$$

$$U_t = 2\phi\phi_t, \tag{18f}$$

$$\mathbf{u}_t + \mathbf{u}\cdot\nabla\mathbf{u} + \nabla p - \frac{1}{\text{Re}}\Delta\mathbf{u} + \frac{1}{\text{We}}(\phi\nabla w_\phi + \psi\nabla w_\psi) = 0, \tag{18g}$$

$$\nabla \cdot \mathbf{u} = 0, \tag{18h}$$

with periodic boundary conditions or

$$\partial_\mathbf{n}\phi^{n+1} = \partial_\mathbf{n}\psi^{n+1} = \partial_\mathbf{n}w_\phi^{n+1} = \partial_\mathbf{n}w_\psi^{n+1} = \mathbf{u}^{n+1} = \partial_\mathbf{n}p^{n+1} = 0, \quad \text{on } \Gamma,$$

where $H(\psi) = G'(\psi)/\sqrt{G(\psi)+B}$.

we can derive the PDE energy law for the transformed governing system (18).

**Theorem 3.1**. The transformed governing system (18) satisfies the following energy dissipation law:

$$\frac{d}{dt} E_{tot}(\phi, \psi, U, V) = -\frac{1}{\text{Pe}_\phi} \int |\nabla w_\phi|^2 \, d\Omega - \frac{1}{\text{Pe}_\psi} \int \left|\sqrt{M_\psi} \nabla w_\psi\right|^2 d\Omega - \text{CaCn} \int |\nabla \mathbf{u}|^2 \, d\Omega \leq 0. \quad (19)$$

*Proof.* By taking the inner product of equation (18a) with $w_\psi$, we obtain,

$$(\psi_t, w_\psi) - (\psi \mathbf{u}, \nabla w_\psi) = -\frac{1}{\text{Pe}_\psi} \int \left|\sqrt{M_\psi} \nabla w_\psi\right|^2 d\Omega, \quad (20)$$

where $(\cdot, \cdot)$ is the inner product in $L^2(\Omega)$.

By taking the inner product of equation (18b) with $-\psi_t$, we can derive that

$$-(w_\psi, \psi_t) = -\text{Pi}(HV, \psi_t) - \frac{1}{4\text{Ex}}(\phi^2, \psi_t) + \frac{1}{4}(U^2, \psi_t). \quad (21)$$

By taking the inner product of equation (18c) with $2\text{Pi}V$, we obtain

$$(V_t, 2\text{Pi}V) = \text{Pi}\frac{d}{dt}\int V^2 \, d\Omega = \text{Pi}(HV, \psi_t). \quad (22)$$

Summing up equations (20) – (22), we obtain

$$\text{Pi}\frac{d}{dt}\int V^2 \, d\Omega - (\psi \mathbf{u}, \nabla w_\psi) = -\frac{1}{4\text{Ex}}(\phi^2, \psi_t) + \frac{1}{4}(U^2, \psi_t) - \frac{1}{\text{Pe}_\psi}\int\left|\sqrt{M_\psi}\nabla w_\psi\right|^2 d\Omega. \quad (23)$$

By taking the inner product of equation (18d) with $w_\phi$, we obtain

$$(\phi_t, w_\phi) - (\mathbf{u}\phi, \nabla w_\phi) = -\frac{1}{\text{Pe}_\phi}\int|\nabla w_\phi|^2 \, d\Omega. \quad (24)$$

By taking the inner product of equation (18e) with $-\phi_t$, we can derive that

$$\begin{aligned}-(w_\phi, \phi_t) &= \frac{\text{Cn}^2}{2}(\Delta\phi, \phi_t) - (\phi U, \phi_t) - \frac{1}{2\text{Ex}}(\psi\phi, \phi_t) + (\psi U\phi, \phi_t) \\ &= -\frac{\text{Cn}^2}{4}\frac{d}{dt}\int|\nabla\phi|^2 \, d\Omega - (\phi U, \phi_t) - \frac{1}{4\text{Ex}}\frac{d}{dt}\int\psi\phi^2 \, d\Omega + \frac{1}{4\text{Ex}}(\psi_t, \phi^2) \\ &\quad + \frac{1}{4}\frac{d}{dt}\int\psi U^2 \, d\Omega - \frac{1}{4}(\psi_t, U^2).\end{aligned} \quad (25)$$

By taking the inner product of equation (18f) with $U/2$, we obtain

$$\frac{1}{2}(U_t, U) = \frac{1}{4}\frac{d}{dt}\int U^2 \, d\Omega = (\phi U, \phi_t). \quad (26)$$

summing up equations (23) – (26), we obtain

$$\begin{aligned}\frac{d}{dt}\int\left(\frac{\text{Cn}^2}{4}|\nabla\phi|^2 + \frac{U^2}{4} + \text{Pi}V^2 + \frac{1}{4\text{Ex}}\psi\phi^2 - \frac{\psi U^2}{4}\right)d\Omega \\ = (\psi \mathbf{u}, \nabla w_\psi) + (\mathbf{u}\phi, \nabla w_\phi) - \frac{1}{\text{Pe}_\psi}\int\left|\sqrt{M_\psi}\nabla w_\psi\right|^2 d\Omega - \frac{1}{\text{Pe}_\phi}\int|\nabla w_\phi|^2 \, d\Omega.\end{aligned} \quad (27)$$

By taking the inner product of equation (18g) with $\text{We}\,\mathbf{u}$, we obtain

$$\frac{\text{We}}{2}\frac{d}{dt}\int|\mathbf{u}|^2 \, d\Omega + \text{CaCn}\int|\nabla\mathbf{u}|^2 \, d\Omega - \text{We}(p, \nabla\cdot\mathbf{u}) + (\phi\mathbf{u}, \nabla w_\phi) + (\psi\mathbf{u}, \nabla w_\psi) = 0. \quad (28)$$

By taking the inner product of equation (18h) with $\text{We}\,p$, we can easily obtain

$$\text{We}(p, \nabla\cdot\mathbf{u}) = 0. \quad (29)$$

Summing up equations (27) – (29), we get the desired results (19)  □

With the introduction of two auxiliary variables $U$ and $V$, the hydrodynamics coupled phase-field surfactant model is transformed into an equivalent form. It can be observed from equations (10) and (19) that the transformed model satisfies the exactly same energy dissipation law with the original system for the time-continuous case. In the next section, we will focus on constructing time-marching schemes with energy stability for the transformed governing system.

**3.2. Linear, decoupled and energy stable schemes**

**3.2.1. First-order scheme**

For phase-field models, constructing efficient and easy-to-implement numerical schemes with energy stability is the main challenge in the numerical approximation [12, 52, 53]. Numerical schemes that do not respect energy dissipation laws may be "overloaded" with an excessive amount of dissipation near singularities, which in turn lead to large numerical errors, particularly for long time integration [12, 54]. Hence, to accurately simulate the interfacial dynamics with surfactants, it is especially desirable to design schemes that satisfy discrete energy laws. In this section, we present a first-order scheme with unconditional energy stability to solve the governing system (18).

Given $\psi^n$, $\phi^n$, $V^n$, $U^n$, $\mathbf{u}^n$ and $p^n$, the scheme (30) calculates $\psi^{n+1}$, $\phi^{n+1}$, $V^{n+1}$, $U^{n+1}$, $\mathbf{u}^{n+1}$ and $p^{n+1}$ for $n \geq 0$ in four steps.

**Step 1,** we update $\psi^{n+1}$ and $V^{n+1}$ by solving

$$\frac{\psi^{n+1} - \psi^n}{\delta t} + \nabla \cdot \left( \mathbf{u}_*^n \psi^n \right) - \frac{1}{\text{Pe}_\psi} \nabla \cdot M_\psi^n \nabla w_\psi^{n+1} = 0, \tag{30a}$$

$$w_\psi^{n+1} = \text{Pi} H^n V^{n+1} + \frac{\left( \phi^n \right)^2}{4\text{Ex}} - \frac{\left( U^n \right)^2}{4}, \tag{30b}$$

$$V^{n+1} - V^n = \frac{1}{2} H^n \left( \psi^{n+1} - \psi^n \right), \tag{30c}$$

$$\mathbf{u}_*^n = \mathbf{u}^n - \frac{\delta t \psi^n}{\text{We}} \nabla w_\psi^{n+1}, \tag{30d}$$

$$\partial_\mathbf{n} \psi^{n+1} = 0, \ \partial_\mathbf{n} w_\psi^{n+1} = 0, \quad \text{on } \Gamma.$$

**Step 2,** update $\phi^{n+1}$ and $U^{n+1}$ using

$$\frac{\phi^{n+1} - \phi^n}{\delta t} + \nabla \cdot \left( \mathbf{u}_{**}^n \phi^n \right) - \frac{1}{\text{Pe}_\phi} \Delta w_\phi^{n+1} = 0, \tag{30e}$$

$$w_\phi^{n+1} = -\frac{\text{Cn}^2}{2} \Delta \phi^{n+1} + \phi^n U^{n+1} + \frac{\psi^{n+1} \phi^{n+1}}{2\text{Ex}} - \frac{1}{2} \psi^{n+1} U^n \left( \phi^{n+1} + \phi^n \right), \tag{30f}$$

$$U^{n+1} - U^n = 2\phi^n \left( \phi^{n+1} - \phi^n \right), \tag{30g}$$

$$\mathbf{u}_{**}^n = \mathbf{u}_*^n - \frac{\delta t \phi^n}{\text{We}} \nabla w_\phi^{n+1}, \tag{30h}$$

$$\partial_\mathbf{n} \phi^{n+1} = 0, \ \partial_\mathbf{n} w_\phi^{n+1} = 0, \quad \text{on } \Gamma.$$

**Step 3,** we update $\tilde{\mathbf{u}}^{n+1}$ by solving

$$\begin{cases} \dfrac{\tilde{\mathbf{u}}^{n+1}-\mathbf{u}^n}{\delta t}-\dfrac{1}{\text{Re}}\Delta\tilde{\mathbf{u}}^{n+1}+\nabla p^n+\left(\mathbf{u}^n\cdot\nabla\right)\tilde{\mathbf{u}}^{n+1}+\dfrac{1}{\text{We}}\left(\phi^n\nabla w_\phi^{n+1}+\psi^n\nabla w_\psi^{n+1}\right)=0,\\ \tilde{\mathbf{u}}^{n+1}=0,\quad\text{on }\Gamma. \end{cases} \qquad (30\text{i})$$

**Step 4,** update $p^{n+1}$ and $\mathbf{u}^{n+1}$ using [12]

$$\begin{cases} \dfrac{\mathbf{u}^{n+1}-\tilde{\mathbf{u}}^{n+1}}{\delta t}+\nabla\left(p^{n+1}-p^n\right)=0,\\ \nabla\cdot\mathbf{u}^{n+1}=0,\\ \mathbf{u}^{n+1}\cdot\mathbf{n}=0,\quad\text{on }\Gamma. \end{cases} \qquad (30\text{j})$$

**Remark 3.1.** (1) Computations of phase-field variables $\psi^{n+1}$ and $\phi^{n+1}$, velocity $\mathbf{u}^{n+1}$ and pressure $p^{n+1}$ are fully decoupled. Moreover, the scheme (30) involves solving only a sequence of linear elliptic equations at each time step, indicating that it is efficient and easy-to-implement. (2) Once we obtained $\psi^{n+1}$ and $\phi^{n+1}$, then $V^{n+1}$ and $U^{n+1}$ can be explicitly calculated from equations (30c) and (30g), respectively. Thus, the introduction of two auxiliary variables $V$ and $U$ do not involve extra computation costs. (3) Following the idea in [12, 55], two first-order stabilization terms are introduced in explicit convective velocities $\mathbf{u}_*^n$ and $\mathbf{u}_{**}^n$, respectively. (4) A logarithmic potential is applied to the phase-field variable $\psi$. At the level of numerical scheme design, positivity-preserving property of $\psi$ is very challenging due to the particularities of the temporal discretization involved [56]. In this study, we use a regularized logarithmic potential [51] in equation (17) to deal with some extreme values of $\psi$. (5) The step 4 can also be rewritten as

$$\begin{cases} -\Delta\left(p^{n+1}-p^n\right)=-\dfrac{1}{\delta t}\nabla\cdot\tilde{\mathbf{u}}^{n+1},\\ \partial_{\mathbf{n}}\left(p^{n+1}-p^n\right)=0,\quad\text{on }\Gamma,\\ \mathbf{u}^{n+1}=\tilde{\mathbf{u}}^{n+1}-\delta t\nabla\left(p^{n+1}-p^n\right). \end{cases}$$

Now, we carry out the energy estimate for the above scheme.

**Theorem 3.2.** The linear, decoupled scheme (30) is unconditionally energy stable, and satisfies the following discrete energy law:

$$E_{tot}^{n+1}-E_{tot}^n \leq -\dfrac{\delta t}{\text{Pe}_\psi}\left\|\sqrt{M_\psi^n}\nabla w_\psi^{n+1}\right\|^2-\dfrac{\delta t}{\text{Pe}_\phi}\left\|\nabla w_\phi^{n+1}\right\|^2-\delta t\text{CaCn}\left\|\nabla\tilde{\mathbf{u}}^{n+1}\right\|^2 \leq 0, \qquad (31)$$

where

$$E_{tot}^n = \dfrac{\text{We}}{2}\|\mathbf{u}^n\|^2+\dfrac{\delta t^2\text{We}}{2}\|\nabla p^n\|^2+\dfrac{\text{Cn}^2}{4}\|\nabla\phi^n\|^2+\dfrac{1}{4}\|U^n\|^2+\text{Pi}\|V^n\|^2 \\ +\dfrac{1}{4\text{Ex}}\left(\psi^n,|\phi^n|^2\right)-\dfrac{1}{4}\left(\psi^n,|U^n|^2\right)-\text{Pi}B|\Omega|, \qquad (32)$$

where $\|\cdot\|$ denotes the $L^2$-norm in $\Omega$.

***Proof.*** By taking the inner product of (30a) with $\delta t w_\psi^{n+1}$, we can easily obtain

$$\left(\psi^{n+1}-\psi^n,w_\psi^{n+1}\right)-\delta t\left(\mathbf{u}_*^n\psi^n,\nabla w_\psi^{n+1}\right)=-\dfrac{\delta t}{\text{Pe}_\psi}\left\|\sqrt{M_\psi^n}\nabla w_\psi^{n+1}\right\|^2. \qquad (33)$$

By taking the inner product of (30b) with $-(\psi^{n+1}-\psi^n)$, we can obtain

$$-\left(\psi^{n+1}-\psi^{n}, w_{\psi}^{n+1}\right)=-\text{Pi}\left(H^{n}V^{n+1}, \psi^{n+1}-\psi^{n}\right)-\frac{1}{4\text{Ex}}\left(\left|\phi^{n}\right|^{2}, \psi^{n+1}-\psi^{n}\right) \\ +\frac{1}{4}\left(\left|U^{n}\right|^{2}, \psi^{n+1}-\psi^{n}\right). \tag{34}$$

Taking the inner product of (30c) with $2\text{Pi}V^{n+1}$ to obtain

$$\text{Pi}\left(\left\|V^{n+1}\right\|^{2}-\left\|V^{n}\right\|^{2}+\left\|V^{n+1}-V^{n}\right\|^{2}\right)=\text{Pi}\left(H^{n}V^{n+1}, \psi^{n+1}-\psi^{n}\right). \tag{35}$$

Summing up equations (33) – (35), we obtain

$$\text{Pi}\left(\left\|V^{n+1}\right\|^{2}-\left\|V^{n}\right\|^{2}+\left\|V^{n+1}-V^{n}\right\|^{2}\right)=\delta t\left(\mathbf{u}_{*}^{n}\psi^{n}, \nabla w_{\psi}^{n+1}\right)-\frac{\delta t}{\text{Pe}_{\psi}}\left\|\sqrt{M_{\psi}^{n}}\nabla w_{\psi}^{n+1}\right\|^{2} \\ -\frac{1}{4\text{Ex}}\left(\left|\phi^{n}\right|^{2}, \psi^{n+1}-\psi^{n}\right)+\frac{1}{4}\left(\left|U^{n}\right|^{2}, \psi^{n+1}-\psi^{n}\right). \tag{36}$$

By taking the inner product of (30e) with $\delta t w_{\phi}^{n+1}$, we obtain

$$\left(\phi^{n+1}-\phi^{n}, w_{\phi}^{n+1}\right)-\delta t\left(\mathbf{u}_{**}^{n}\phi^{n}, \nabla w_{\phi}^{n+1}\right)=-\frac{\delta t}{\text{Pe}_{\phi}}\left\|\nabla w_{\phi}^{n+1}\right\|^{2}. \tag{37}$$

By taking the inner product of (30f) with $-(\phi^{n+1}-\phi^{n})$, we can obtain

$$-\left(\phi^{n+1}-\phi^{n}, w_{\phi}^{n+1}\right)=-\frac{\text{Cn}^{2}}{2}\left(\nabla\phi^{n+1}, \nabla\phi^{n+1}-\nabla\phi^{n}\right)-\left(U^{n+1}\phi^{n}, \phi^{n+1}-\phi^{n}\right) \\ -\frac{1}{2\text{Ex}}\left(\psi^{n+1}\phi^{n+1}, \phi^{n+1}-\phi^{n}\right)+\frac{1}{2}\left(\psi^{n+1}U^{n}\left(\phi^{n+1}+\phi^{n}\right), \phi^{n+1}-\phi^{n}\right) \\ =-\frac{\text{Cn}^{2}}{4}\left(\left\|\nabla\phi^{n+1}\right\|^{2}-\left\|\nabla\phi^{n}\right\|^{2}+\left\|\nabla\phi^{n+1}-\nabla\phi^{n}\right\|^{2}\right)-\left(U^{n+1}\phi^{n}, \phi^{n+1}-\phi^{n}\right) \\ -\frac{1}{4\text{Ex}}\left[\left(\psi^{n+1}, \left|\phi^{n+1}\right|^{2}\right)-\left(\psi^{n+1}, \left|\phi^{n}\right|^{2}\right)+\left(\psi^{n+1}, \left|\phi^{n+1}-\phi^{n}\right|^{2}\right)\right] \\ +\frac{1}{4}\left[\left(\psi^{n+1}, \left|U^{n+1}\right|^{2}\right)-\left(\psi^{n+1}, \left|U^{n}\right|^{2}\right)-\left(\psi^{n+1}, \left|U^{n+1}-U^{n}\right|^{2}\right)\right]. \tag{38}$$

where $\psi^{n}$ is the surfactant concentration at the $n^{th}$ step, which is between 0 and 1.

Taking the inner product of (30g) with $U^{n+1}/2$, we obtain

$$\frac{1}{4}\left(\left\|U^{n+1}\right\|^{2}-\left\|U^{n}\right\|^{2}+\left\|U^{n+1}-U^{n}\right\|^{2}\right)=\left(U^{n+1}\phi^{n}, \phi^{n+1}-\phi^{n}\right). \tag{39}$$

Summing up equations (36) – (39), and dropping off some positive terms, we obtain

$$\frac{\text{Cn}^{2}}{4}\left(\left\|\nabla\phi^{n+1}\right\|^{2}-\left\|\nabla\phi^{n}\right\|^{2}\right)+\frac{1}{4}\left(\left\|U^{n+1}\right\|^{2}-\left\|U^{n}\right\|^{2}\right)+\text{Pi}\left(\left\|V^{n+1}\right\|^{2}-\left\|V^{n}\right\|^{2}\right) \\ +\frac{1}{4\text{Ex}}\left[\left(\psi^{n+1}, \left|\phi^{n+1}\right|^{2}\right)-\left(\psi^{n}, \left|\phi^{n}\right|^{2}\right)\right]-\frac{1}{4}\left[\left(\psi^{n+1}, \left|U^{n+1}\right|^{2}\right)-\left(\psi^{n}, \left|U^{n}\right|^{2}\right)\right] \\ \leq-\frac{\delta t}{\text{Pe}_{\psi}}\left\|\sqrt{M_{\psi}^{n}}\nabla w_{\psi}^{n+1}\right\|^{2}-\frac{\delta t}{\text{Pe}_{\phi}}\left\|\nabla w_{\phi}^{n+1}\right\|^{2}+\delta t\left(\mathbf{u}_{*}^{n}\psi^{n}, \nabla w_{\psi}^{n+1}\right)+\delta t\left(\mathbf{u}_{**}^{n}\phi^{n}, \nabla w_{\phi}^{n+1}\right). \tag{40}$$

We can derive from equations (30d) and (30h) that

$$\text{We}\left(\tilde{\mathbf{u}}^{n+1}-\mathbf{u}^{n}\right)+\delta t\left(\psi^{n}\nabla w_{\psi}^{n+1}+\phi^{n}\nabla w_{\phi}^{n+1}\right)=\text{We}\left(\tilde{\mathbf{u}}^{n+1}-\mathbf{u}_{**}^{n}\right). \tag{41}$$

Now, by taking the inner product of (30i) with $2\delta t\text{We}\tilde{\mathbf{u}}^{n+1}$, and using equation (41), we can obtain

$$\text{We}\left(\left\|\tilde{\mathbf{u}}^{n+1}\right\|^{2}-\left\|\mathbf{u}_{**}^{n}\right\|^{2}+\left\|\tilde{\mathbf{u}}^{n+1}-\mathbf{u}_{**}^{n}\right\|^{2}\right)+2\delta t\text{CaCn}\left\|\nabla\tilde{\mathbf{u}}^{n+1}\right\|^{2}+2\delta t\text{We}\left(\nabla p^{n}, \tilde{\mathbf{u}}^{n+1}\right)=0. \tag{42}$$

By taking the inner product of (30j) with $2\delta t^{2}\text{We}\nabla p^{n}$ and $\mathbf{u}^{n+1}$ separately, we obtain

$$\delta t^2 \mathrm{We} \left( \left\| \nabla p^{n+1} \right\|^2 - \left\| \nabla p^n \right\|^2 - \left\| \nabla p^{n+1} - \nabla p^n \right\|^2 \right) = 2 \delta t \mathrm{We} \left( \tilde{\mathbf{u}}^{n+1}, \nabla p^n \right), \tag{43}$$

and

$$\mathrm{We} \left( \left\| \mathbf{u}^{n+1} \right\|^2 + \left\| \tilde{\mathbf{u}}^{n+1} - \mathbf{u}^n \right\|^2 \right) = \mathrm{We} \left\| \tilde{\mathbf{u}}^{n+1} \right\|^2. \tag{44}$$

We also derive from (30j) that

$$\delta t^2 \mathrm{We} \left\| \nabla p^{n+1} - \nabla p^n \right\|^2 = \mathrm{We} \left\| \tilde{\mathbf{u}}^{n+1} - \mathbf{u}^n \right\|^2. \tag{45}$$

Combining equations (42) – (45), we obtain

$$\mathrm{We} \left( \left\| \mathbf{u}^{n+1} \right\|^2 - \left\| \mathbf{u}_{**}^n \right\|^2 + \left\| \tilde{\mathbf{u}}^{n+1} - \mathbf{u}_{**}^n \right\|^2 \right) + 2 \delta t \mathrm{CaCn} \left\| \nabla \tilde{\mathbf{u}}^{n+1} \right\|^2 + \delta t^2 \mathrm{We} \left( \left\| \nabla p^{n+1} \right\|^2 - \left\| \nabla p^n \right\|^2 \right) = 0. \tag{46}$$

By taking the inner product of (30d) with $\mathbf{u}_*^n$, we can easily obtain

$$\mathrm{We} \left( \left\| \mathbf{u}_*^n \right\|^2 - \left\| \mathbf{u}^n \right\|^2 + \left\| \mathbf{u}_*^n - \mathbf{u}^n \right\|^2 \right) = -2 \delta t \mathrm{We} \left( \psi^n \nabla w_\psi^{n+1}, \mathbf{u}_*^n \right). \tag{47}$$

Similarly, taking the inner product of (30h) with $\mathbf{u}_{**}^n$ to derive

$$\mathrm{We} \left( \left\| \mathbf{u}_{**}^n \right\|^2 - \left\| \mathbf{u}_*^n \right\|^2 + \left\| \mathbf{u}_{**}^n - \mathbf{u}_*^n \right\|^2 \right) = -2 \delta t \mathrm{We} \left( \phi^n \nabla w_\phi^{n+1}, \mathbf{u}_{**}^n \right). \tag{48}$$

Adding equations (46) – (48) together, and dropping off some positive terms, we obtain

$$\begin{aligned}
\frac{\mathrm{We}}{2} \left( \left\| \mathbf{u}^{n+1} \right\|^2 - \left\| \mathbf{u}^n \right\|^2 \right) + \delta t \mathrm{CaCn} \left\| \nabla \tilde{\mathbf{u}}^{n+1} \right\|^2 + \frac{\delta t^2 \mathrm{We}}{2} \left( \left\| \nabla p^{n+1} \right\|^2 - \left\| \nabla p^n \right\|^2 \right) \\
\leq -\delta t \left( \psi^n \nabla w_\psi^{n+1}, \mathbf{u}_*^n \right) - \delta t \left( \phi^n \nabla w_\phi^{n+1}, \mathbf{u}_{**}^n \right).
\end{aligned} \tag{49}$$

Finally, combining (40) and (49), we arrive at

$$\begin{aligned}
& \frac{\mathrm{We}}{2} \left( \left\| \mathbf{u}^{n+1} \right\|^2 - \left\| \mathbf{u}^n \right\|^2 \right) + \frac{\delta t^2 \mathrm{We}}{2} \left( \left\| \nabla p^{n+1} \right\|^2 - \left\| \nabla p^n \right\|^2 \right) + \frac{\mathrm{Cn}^2}{4} \left( \left\| \nabla \phi^{n+1} \right\|^2 - \left\| \nabla \phi^n \right\|^2 \right) \\
& + \frac{1}{4} \left( \left\| U^{n+1} \right\|^2 - \left\| U^n \right\|^2 \right) + \mathrm{Pi} \left( \left\| V^{n+1} \right\|^2 - \left\| V^n \right\|^2 \right) \\
& + \frac{1}{4 \mathrm{Ex}} \left[ \left( \psi^{n+1}, \left| \phi^{n+1} \right|^2 \right) - \left( \psi^n, \left| \phi^n \right|^2 \right) \right] - \frac{1}{4} \left[ \left( \psi^{n+1}, \left| U^{n+1} \right|^2 \right) - \left( \psi^n, \left| U^n \right|^2 \right) \right] \\
& \leq -\frac{\delta t}{\mathrm{Pe}_\psi} \left\| \sqrt{M_\psi^n} \nabla w_\psi^{n+1} \right\|^2 - \frac{\delta t}{\mathrm{Pe}_\phi} \left\| \nabla w_\phi^{n+1} \right\|^2 - \delta t \mathrm{CaCn} \left\| \nabla \tilde{\mathbf{u}}^{n+1} \right\|^2 \leq 0,
\end{aligned} \tag{50}$$

which implies the desired results (31). □

**Remark 3.2.** (1) By using the IEQ approach, for the first time, we proposed a liner, totally decoupled and unconditionally energy stable scheme for a hydrodynamics coupled phase-field surfactant model. Unlike traditional approaches, such as convex splitting and stabilization approaches, the IEQ approach only requires the nonlinear parts of free energy potential are bounded from below, and it is not restricted to the specific forms of nonlinear parts [3]. In this study, it works well for the fourth-order double well potential and logarithmic Flory-Huggins potential. With the introduction of two auxiliary variables $U$ and $V$, nonlinear potentials are transformed into quadratic forms, which provides the fundamental support for the linearization method. (2) At the PDE level, the original energy functional and reformulated energy functional are equivalent and satisfy the exactly same energy dissipation law. However, at the numerical level, the reformulated energy functional in equation (32) is different from the original energy functional due to the introduction of the term $\delta t^2 \text{We} \|\nabla p^n\|^2 / 2$. Thus, the discrete energy dissipation law for the reformulated energy functional may not be available for the original energy functional. (3) Convergence analysis and error estimates are probably available for the proposed time-marching scheme, and some pioneering works for phase-field models with fluid flow can refer to [44, 57-59].

### 3.2.2. Second-order scheme

Schemes with unconditional energy stability remove constraints on the time step-size from the stability point of view. Nonetheless, we should notice that larger time step will definitely introduce larger numerical errors. To use the time step-size as large as possible while maintaining the desirable accuracy, more accurate and energy sable schemes are needed, e.g., second-order schemes. In this section, we extend the above first-order scheme to a second-order version based on the backward differentiation formula [46].

Given $\psi^{n-1}, \phi^{n-1}, p^{n-1}, U^{n-1}, V^{n-1}, \psi^n, \phi^n, p^n, V^n$, and $U^n$, the scheme calculates $\psi^{n+1}, \phi^{n+1}, p^{n+1}, U^{n+1}$ and $V^{n+1}$ in two steps.

**Step 1**, update $\psi^{n+1}$ and $V^{n+1}$ using,

$$\begin{cases} \dfrac{3\psi^{n+1} - 4\psi^n + \psi^{n-1}}{2\delta t} + \nabla \cdot \left(\mathbf{u}_*^n \psi^*\right) = \dfrac{1}{\text{Pe}_\psi} \nabla \cdot \left(M_\psi^* \nabla w_\psi^{n+1}\right), \\ w_\psi^{n+1} = \text{Pi} H\left(\psi^*\right) V^{n+1} + \dfrac{\left(\phi^*\right)^2}{4\text{Ex}} - \dfrac{\left(U^*\right)^2}{4}, \\ \dfrac{3V^{n+1} - 4V^n + V^{n-1}}{2\delta t} = \dfrac{1}{2} H\left(\psi^*\right) \left(\dfrac{3\psi^{n+1} - 4\psi^n + \psi^{n-1}}{2\delta t}\right), \\ \mathbf{u}_*^n = \mathbf{u}^* - \dfrac{\delta t \psi^*}{\text{We}} \nabla w_\psi^{n+1}, \\ \partial_\mathbf{n} \psi^{n+1} = \partial_\mathbf{n} w_\psi^{n+1} = 0, \quad \text{on } \Gamma. \end{cases} \quad (51a)$$

**Step 2**, we update $\phi^{n+1}$ and $U^{n+1}$ by solving

$$\begin{cases} \dfrac{3\phi^{n+1} - 4\phi^n + \phi^{n-1}}{2\delta t} + \nabla \cdot \left(\mathbf{u}_{**}^n \phi^*\right) = \dfrac{1}{\text{Pe}_\phi} \Delta w_\phi^{n+1}, \\ w_\phi^{n+1} = -\dfrac{\text{Cn}^2}{2} \Delta \phi^{n+1} + \phi^* U^{n+1} + \dfrac{\psi^{n+1} \phi^{n+1}}{2\text{Ex}} - \psi^{n+1} U^* \phi^{n+1}, \\ \dfrac{3U^{n+1} - 4U^n + U^{n-1}}{2\delta t} = 2\phi^* \left(\dfrac{3\phi^{n+1} - 4\phi^n + \phi^{n-1}}{2\delta t}\right), \\ \mathbf{u}_{**}^n = \mathbf{u}_*^n - \dfrac{\delta t \phi^*}{\text{We}} \nabla w_\phi^{n+1}, \\ \partial_\mathbf{n} \phi^{n+1} = \partial_\mathbf{n} w_\phi^{n+1} = 0, \quad \text{on } \Gamma. \end{cases} \quad (51b)$$

**Step 3**, we update $\tilde{\mathbf{u}}^{n+1}$ by solving

$$\begin{cases} \dfrac{3\tilde{\mathbf{u}}^{n+1} - 4\mathbf{u}^n + \mathbf{u}^{n-1}}{2\delta t} + \mathbf{u}^* \cdot \nabla \tilde{\mathbf{u}}^{n+1} - \dfrac{1}{\text{Re}} \Delta \tilde{\mathbf{u}}^{n+1} + \nabla p^n + \dfrac{1}{\text{We}} \left(\phi^* \nabla w_\phi^{n+1} + \psi^* \nabla w_\psi^{n+1}\right) = 0, \\ \tilde{\mathbf{u}}^{n+1} = 0, \quad \text{on } \Gamma. \end{cases} \quad (51c)$$

**Step 4**, update $p^{n+1}$ and $\mathbf{u}^{n+1}$ using

$$\begin{cases} \Delta\left(p^{n+1}-p^{n}\right)=\dfrac{3}{2\delta t}\nabla\cdot\tilde{\mathbf{u}}^{n+1}, \\ \mathbf{u}^{n+1}=\tilde{\mathbf{u}}^{n+1}-\dfrac{2\delta t}{3}\nabla\left(p^{n+1}-p^{n}\right), \\ \partial_{\mathbf{n}}\left(p^{n+1}-p^{n}\right)=0,\quad \text{on } \Gamma. \end{cases} \quad (51d)$$

where

$$\begin{cases} \phi^{*}=2\phi^{n}-\phi^{n-1},\quad \psi^{*}=2\psi^{n}-\psi^{n-1},\quad U^{*}=2U^{n}-U^{n-1}, \\ V^{*}=2V^{n}-V^{n-1},\quad \mathbf{u}^{*}=2\mathbf{u}^{n}-\mathbf{u}^{n-1},\quad M_{\psi}^{*}=2M_{\psi}^{n}-M_{\psi}^{n-1}. \end{cases}$$

Nonlinearly coupled terms between phase-field variables present a huge challenge to carry out the energy estimate. Although the unconditional stability of the second-order scheme still needs to be proved, a series of numerical tests performed in Section 4.1 have shown that the proposed second-order scheme is energy stable.

## 4. Numerical results

An efficient finite difference method on staggered grids is used for the spatial discretization. Two phase-field variables $\phi$ and $\psi$, two auxiliary variables $U$, $V$, as well as pressure $p$ are defined at the center of each cell. Velocities in $x$, $y$ and $z$ directions are defined at the cell face centers. The advection terms in the Cahn-Hilliard-type equations and the Navier-Stokes equation are discretized by a composite high resolution scheme. More precisely, the fluxes at cell faces are evaluated with a MINMOD scheme [60, 61]. The MINMOD scheme not only achieves the second-order accuracy, but also preserves the physical properties of convection. Details on the MINMOD scheme can be found in [61]. Other spatial derivatives in schemes (28) and (49) are discretized using the standard central difference scheme. A preconditioned BICGSTAB solver is used to solve the above variables. It is worth emphasizing that $V^{n+1}$ and $U^{n+1}$ do not involve any extra computational cost, since they can be calculated explicitly once we obtain $\psi^{n+1}$ and $\phi^{n+1}$. In this section, we conduct a series of numerical results to demonstrate that the proposed schemes are accurate, efficient and energy stable.

### 4.1. Droplet deformation under a shear flow

We first use the shear flow case to test the accuracy of the proposed schemes. All simulations are conducted in a rectangular domain $[0, 6] \times [0, 4]$ with periodic boundary conditions on the left and right sides, as shown in Fig 1(a). Equal but opposite velocities are prescribed on top and bottom walls, respectively. Initially, a circular bubble with the radius of $R=1$ locates in the center of the domain. The surfactant bulk concentration $\psi_b$ is $1\times10^{-4}$. The bubble radius $R$ and top wall moving velocity are chosen as the characteristic length and velocity, respectively. For all experiments in Section 4, Pi and $\psi_c$ take 0.1227 and 0.017, respectively, and Ex is determined by equation (16). Other simulation parameters are as follows:

$$\text{Pe}_\phi = 10,\quad \text{Pe}_\psi = 100,\quad \text{Re} = 0.5,\quad \text{Ca} = 0.5,\quad \text{Cn} = 0.025 \text{ and } \xi = 1\times10^{-7}.$$

LS1 and LS2 are used to represent the first-order scheme and second-order scheme, respectively. We use the LS2 with a small time step-size $\delta t = 6.25\times10^{-5}$ to calculate the reference solutions due to the lack of exact solutions. To reduce the error related to spatial discretization as much as possible, a spatial resolution $n_x = 324$, $n_y = 216$ is used in simulations. A series of time step-sizes $\delta t = 2\times10^{-3}$, $1\times10^{-3}$, $5\times10^{-4}$, $2.5\times10^{-4}$ and $1.25\times10^{-4}$ are chosen to calculate the $L^2$-norm errors and convergence orders for phase-field variables $\phi$ and $\psi$ at $t = 0.5$. Table 1 demonstrates that LS1 and LS2 schemes can achieve the first-order and second-order accuracy in time, respectively.

Moreover, the second-order scheme LS2 gives better accuracy than the first-order scheme at the same time step-size.

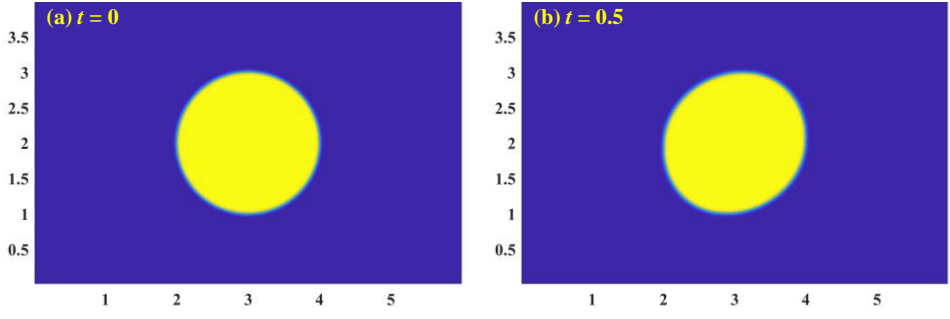

Fig 1 Time evolution of a droplet under a shear flow. ($\psi_b = 1\times10^{-4}$)

**Table 1.** The $L^2$ norm errors and convergence order for phase-field variables $\phi$ and $\psi$ at $t = 0.5$ with different temporal resolutions.

| $\Delta t$ | LS1_ph | Order | LS1_psi | Order | LS2_ph | Order | LS2_psi | Order |
|---|---|---|---|---|---|---|---|---|
| $2\times10^{-3}$ | $4.14\times10^{-2}$ | - | $1.73\times10^{-4}$ | - | $4.40\times10^{-3}$ | - | $6.53\times10^{-5}$ | - |
| $1\times10^{-3}$ | $2.37\times10^{-2}$ | 0.81 | $1.05\times10^{-4}$ | 0.72 | $1.32\times10^{-3}$ | 1.74 | $1.88\times10^{-5}$ | 1.82 |
| $5\times10^{-4}$ | $1.21\times10^{-2}$ | 0.98 | $5.45\times10^{-5}$ | 0.95 | $3.52\times10^{-4}$ | 1.91 | $4.92\times10^{-6}$ | 1.91 |
| $2.5\times10^{-4}$ | $5.39\times10^{-3}$ | 1.16 | $2.45\times10^{-5}$ | 1.15 | $8.51\times10^{-5}$ | 2.05 | $1.22\times10^{-6}$ | 1.98 |
| $1.25\times10^{-4}$ | $1.83\times10^{-3}$ | 1.56 | $8.33\times10^{-5}$ | 1.56 | $2.12\times10^{-5}$ | 2.01 | $2.81\times10^{-7}$ | 2.07 |

Now we study the energy stability of the proposed schemes (LS1 and LS2). To ensure no input energy from the outside, velocities of bottom and top walls $u_w$ are set to zero. We take the shape of droplet at $t = 2.5$ (Fig 3(b)) as the initial state. The droplet will gradually turn into a circle under the effect of capillary force. Evolution of total energy in Fig 2 confirms the energy stability of the proposed schemes. Considerable differences between different time step-sizes can also be observed, indicating that the induced numerical errors with large time step-sizes are higher.

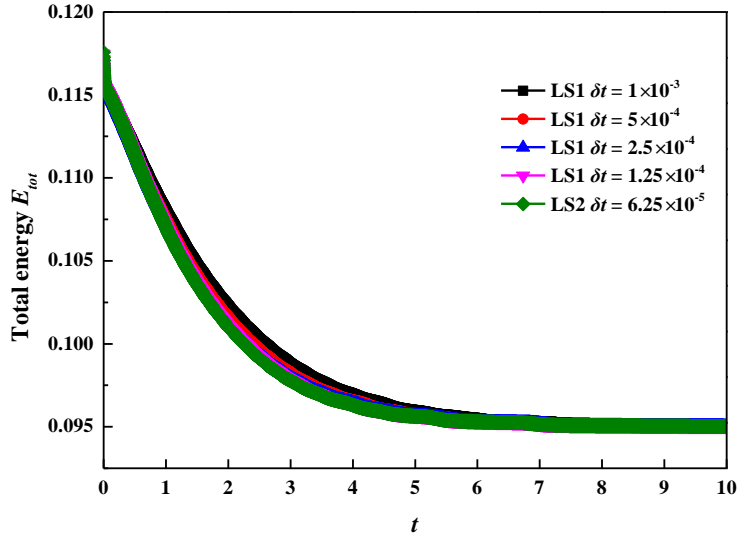

Fig 2 Energy curves at different time step-sizes.

To examine the effect of surfactants, we consider three different surfactant bulk concentrations ($\psi_b = 1\times10^{-4}$, $5\times10^{-3}$ and $1.5\times10^{-2}$). We use the second-order scheme (LS2) with grid size $324\times216$ and time step-size $\delta t = 5 \times 10^{-4}$ to simulate the droplet deformation under a shear flow. The aforementioned parameters will continue to be used in simulations. We run each simulation until $t = 20$. Fig 3 gives the evolution of a droplet under a shear flow with the presence of surfactants ($\psi_b = 1.5 \times 10^{-2}$). It can be observed that surfactants gradually migrate towards to tip ends under a shear flow. The movement of surfactants causes uneven distribution of interfacial tension, and the smallest interfacial tension occurs at the droplet tips. The non-uniform distribution of surfactants around the interface will arise Marangoni force, which in turn prevents the migration of surfactants.

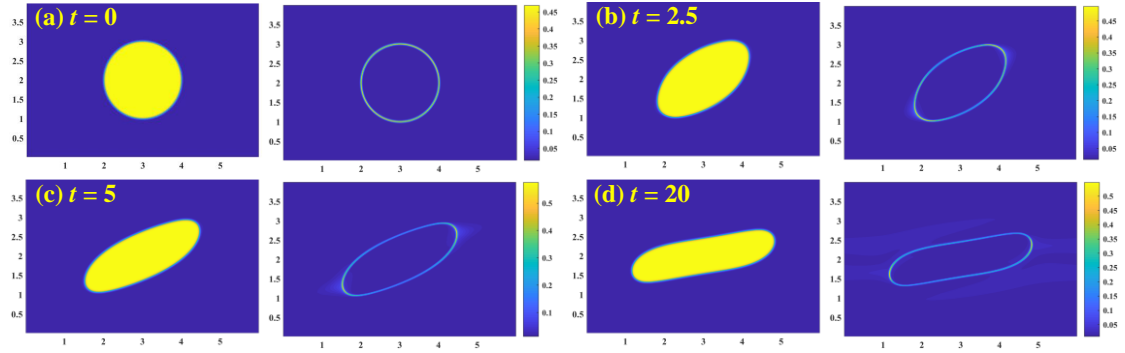

Fig 3 Time evolution of a droplet and surfactant concentration in a shear flow ($\psi_b = 1.5 \times 10^{-2}$). For each subfigure, the left is the profile of $\phi$, and the right is the profile of $\psi$.

In Fig 4, we depict droplet profiles at different surfactant bulk concentrations $\psi_b$. As we expected, high surfactant bulk concentration results in a more prolate droplet, and differences between droplet profiles increase with time. These results perfectly reflect the role of surfactants in reducing interfacial tension.

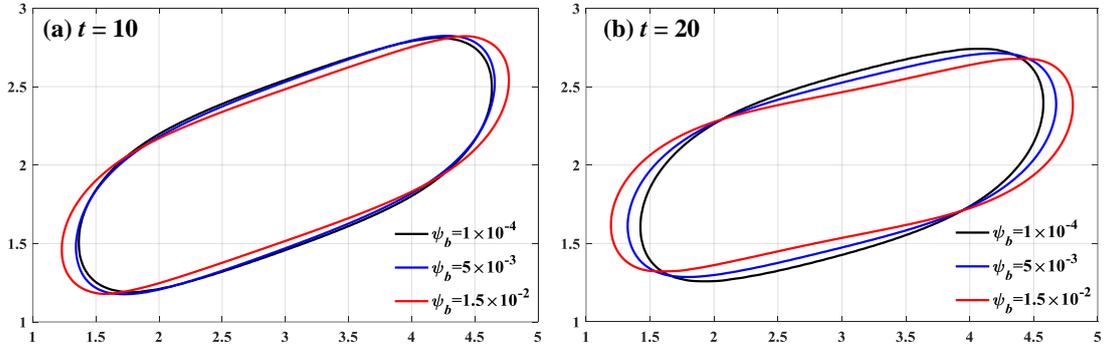

Fig 4 Time evolution of droplets under a shear flow with the presence of surfactants (black line: $\psi_b = 1 \times 10^{-4}$; blue line: $\psi_b = 5 \times 10^{-3}$; red line: $\psi_b = 1.5 \times 10^{-2}$).

We extend the study of droplet deformation to a 3D domain $\Omega = [0, 1] \times [0, 1] \times [0, 1]$. We use a spatial resolution $100^3$ and a time step-size $\delta t = 5 \times 10^{-4}$. $x$ and $y$ directions are periodic boundaries. The side length of the domain and top wall moving velocity are chosen as the characteristic length and velocity, respectively. The first-order scheme (LS1) with energy stability is adopted to perform the simulation. To facilitate the reader to reproduce our results, we list all paramters used in simulations:

$$\text{Pe}_\phi = 10, \quad \text{Pe}_\psi = 500, \quad \text{Re} = 10, \quad \text{Ca} = 0.2, \quad \text{Cn} = 0.015 \text{ and } \xi = 1 \times 10^{-7}.$$

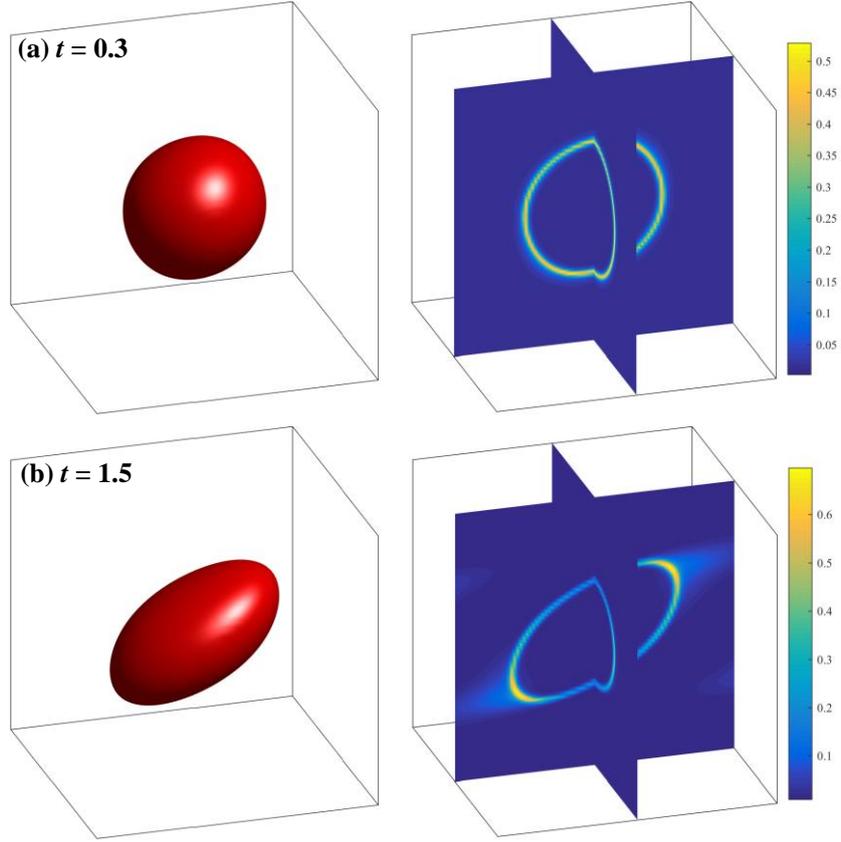

Fig 5 Time evolution of a droplet and surfactant concentration under the effect of shear flow ($\psi_b = 1.5 \times 10^{-2}$). For each subfigure, the left is the profile of $\phi$, and the right is the profile of $\psi$.

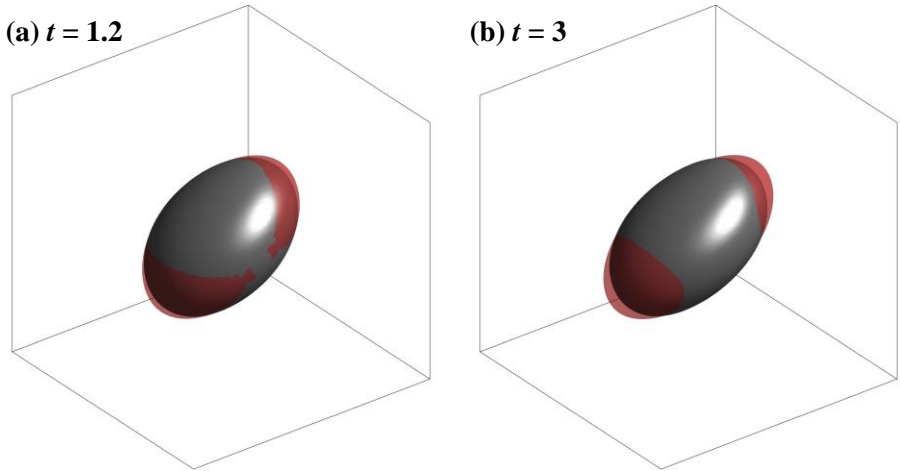

Fig 6 Evolutions of phase-field variable $\phi$ in a shear flow at $t = 1.2$ (left) and $t = 3$ (right). (gray: $\psi_b = 1 \times 10^{-4}$; red: $\psi_b = 1.5 \times 10^{-2}$)

Similar to the 2D results, the droplet continuously deforms under the effect of shear flow, and surfactants gradually swept towards to the tip ends of droplet (Fig 5(a)). Surfactants are convected into the bulk phases when the concentration of surfactants at the tip ends reach the maximum, as shown in Fig 5(b). Comparison of droplet profiles at the different surfactant bulk concentrations is presented in Fig 6. Again, the results demonstrate the role of surfactant in reducing surface tension.

### 4.2. Coarsening Dynamics

To demonstrate the unconditional stability of the proposed numerical scheme, we use a 2D domain $\Omega = [0, 1] \times [0, 1]$ to simulate the coarsening dynamics in the presence of flow. The initial conditions are taken as the randomly perturbed concentration fields: $\phi = 0.1+0.001\times\text{rand}\,(x, y)$, $\rho = 0.01+0.001\times\text{rand}\,(x, y)$. The function rand $(x, y)$ represents the random number in $[0, 1]$. All boundaries are periodic. We use the first-order scheme (LS1) with a grid size of $200^2$ and time step size of $\delta t = 1\times 10^{-4}$ to simulate the coarsening dynamics. Other parameters used in simulations are listed as follows:

$$\text{Pe}_\phi = 100, \quad \text{Pe}_\psi = 100, \quad \text{Re} = 1, \quad \text{Ca} = 1, \quad \text{Cn} = 0.01 \text{ and } \xi = 1\times 10^{-7}.$$

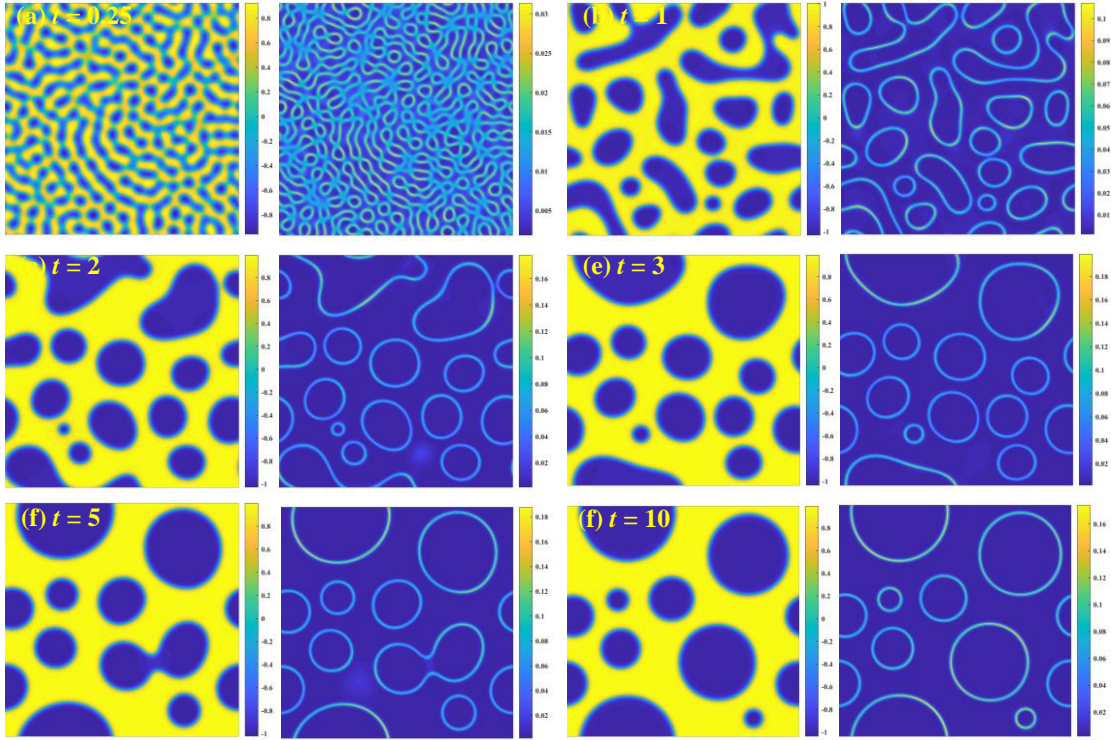

Fig 7 Evolution of the phase-field variables $\phi$ and $\psi$ at different times. For each subfigure, the left is the profile of $\phi$, and the right is the profile of $\psi$.

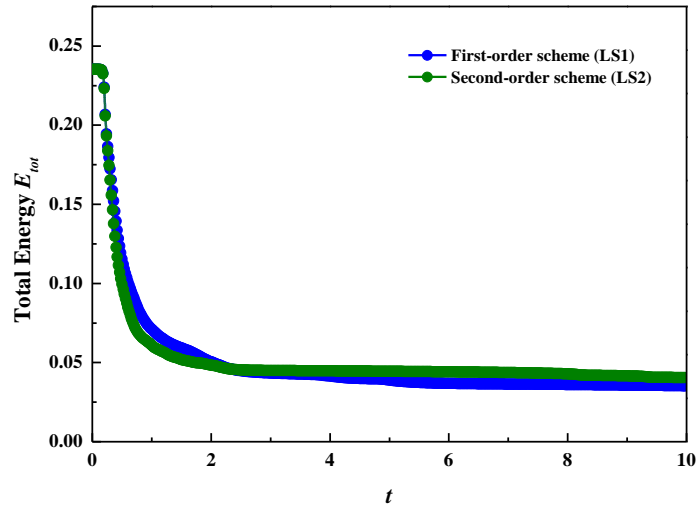

Fig 8 Curves of total energy obtained by the first-order scheme (LS1) and second-order scheme (LS2).

Fig 7 gives the evolutions of phase-field variables $\phi$ and $\psi$ at different times. As we expected, surfactants are automatically absorbed to the interface of fluids. Energy curves in Fig 8 demonstrate that the proposed schemes (LS1 and LS2) are energy stable.

**4.3. Colliding droplets**

In this section, we perform the collision of two equal-sized droplets under a shear flow with the first-order scheme (LS1) to investigate the effect of surfactants on droplet-droplet interactions. Simulations are conducted on a 6 × 4 flow domain with a spatial resolution 360 × 240 and a time step-size $\delta t = 5 \times 10^{-4}$. Initially, two circular droplets with the radius of 0.7 locate at (1.5, 2.5) and (4.5, 1.5), respectively. The left and right sides of the domain are periodic boundary conditions. The detailed simulation parameters are as follows:

$$\text{Pe}_\phi = 10, \quad \text{Pe}_\psi = 100, \quad \text{Re} = 0.5, \quad \text{Ca} = 0.25, \quad \text{Cn} = 0.025 \text{ and } \xi = 1 \times 10^{-7}.$$

The left column of Fig 9 presents the evolution of two equal-sized droplets with high surfactant bulk concentration ($\psi_b = 1.5 \times 10^{-2}$). The high surfactant bulk concentration prevents the droplet coalescence due to the collision, which is consistent with some experimental and numerical observations [8, 34]. The right column of Fig 9 shows the concentration of surfactants. At the beginning, surfactants migrate towards the tips of each droplet. The pressure in the gap between two droplets sees a significant increase when two droplets approach each other, which pushes surfactants away from the near-contact region, as shown in Fig 9(a) and Fig 9(b). Similarly, the uneven distribution of surfactants along the interface will arise the Marangoni force, which acts as an additional repulsive force to prevent droplet coalescence [6, 8]. Another non-negligible fact is that the presence of surfactants can enhance the deformation of droplet, and thus affect the collision process of droplets as well.

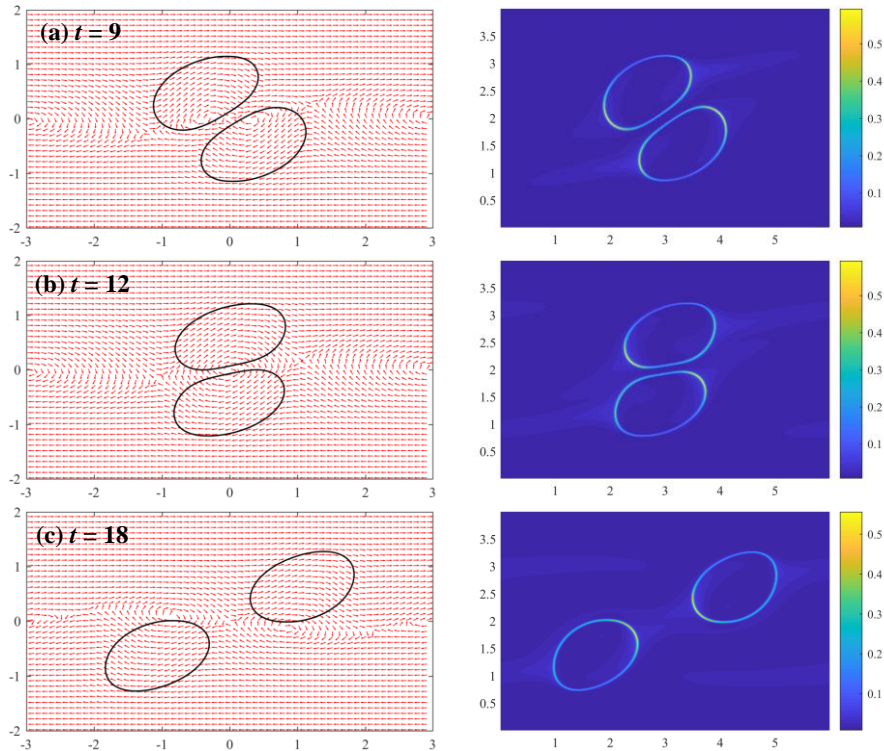

Fig 9 Evolution of two equal-sized droplets and surfactant concentration under a linear shear flow ($\psi_b = 1.5 \times 10^{-2}$). For each subfigure, the left is the profile of $\phi$, and the right is the profile of $\psi$.

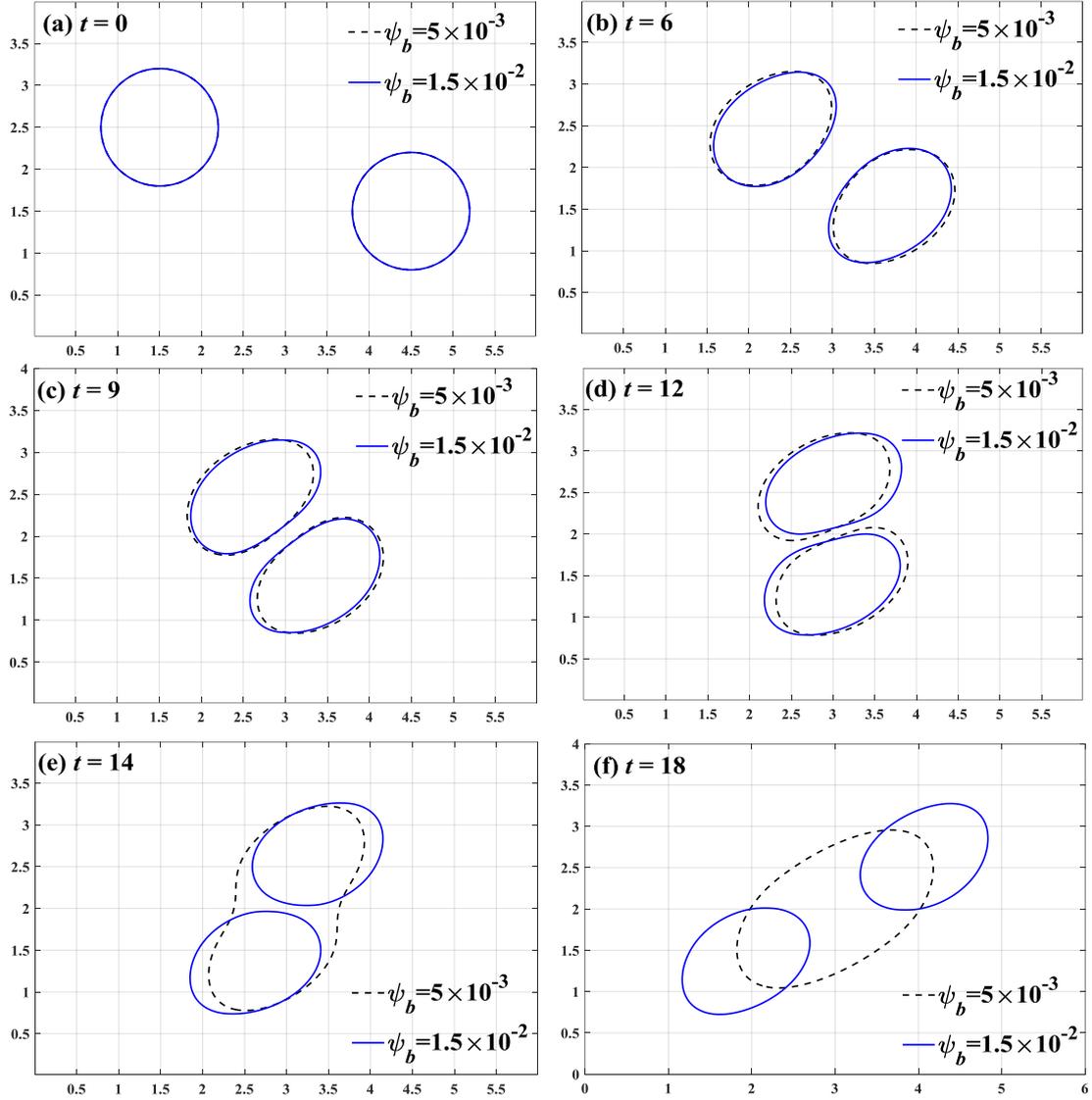

Fig 10 The collision of two droplets in a shear flow. (black dash line: $\psi_b=5\times10^{-3}$; blue solid line: $\psi_b=1.5\times10^{-2}$)

Figure 10 compares the collision of two droplets at different surfactant bulk concentrations under a shear flow. It can be observed that the two droplets collide and merge together at low surfactant concentration.

## 5. Conclusions

In this study, we simulate the interfacial dynamics with soluble surfactants in a multiphase system. An energy-based diffuse interface model for two-phase flows with soluble surfactants is presented. A first and a second-order time marching schemes, which are linear and totally decoupled, for a hydrodynamics coupled phase-field surfactant model are proposed, and the first-order scheme is unconditionally energy stable. Appropriate auxiliary variables are introduced to transform the governing system into an equivalent form, which allows the nonlinear potentials to be treated efficiently and semi-explicitly. At each time step, the schemes involve solving a sequence of linear elliptic equations, and computations of phase-field variables, velocity and pressure are totally decoupled. We also carry out the energy estimate for the first-order semi-implicit scheme. Various 2D and 3D numerical experiments demonstrate that the surfactant concentration has significant impact on the droplet deformation and collision under a shear flow. The increase in surfactant concentration can promote droplet deformation. The uneven distribution of surfactants along the interface will arise the Marangoni force, which acts as an additional repulsive force to prevent droplet coalescence.

**Acknowledgement**

Jun Yao and Guangpu Zhu acknowledge that this work is supported by the National Science and Technology Major Project (2016ZX05011-001), the NSF of China (51804325, 51504276, and 51674280). The work of Shuyu Sun and Jisheng Kou is supported by the KAUST research fund awarded to the Computational Transport Phenomena Laboratory at KAUST through the Grant BAS/1/1351-01-01.